\documentclass[preprint,12pt]{elsarticle}
\usepackage{subfig}

 \usepackage{graphicx}

\usepackage{amssymb}

\journal{Nuclear Physics B}
\newcommand{\newc}{\newcommand}
\newc{\lra}{\leftrightarrow}
\newc{\beq}{\begin{equation}}
\newc{\eeq}{\end{equation}}
\newc{\barr}{\begin{eqnarray}}
\newc{\earr}{\end{eqnarray}}

\begin{document}

\begin{frontmatter}
\date{\today}

\title {Probing the fourth neutrino existence by neutral current oscillometry in the spherical gaseous TPC }


\author{J.D. Vergados$^{1}$, Y. Giomataris$^{2}$ and Yu.N. Novikov$^{3}$}

\address{
1 University of Ioannina, Ioannina, GR 45110, Greece
\\E-mail\footnote{Corresponding author}:Vergados@uoi.gr}
\address{
2 CEA, Saclay, DAPNIA, Gif-sur-Yvette, Cedex,France}
\address{
3 Petersburg Nuclear Physics Institute, 188300, Gatchina,Russia and St.Petersburg State University, 199034 St.Petersburg, Russia }

\begin{abstract}

It is shown that, if  the "new neutrino"  implied by the Reactor Neutrino Anomaly exists and is in fact characterized by the suggested relatively   high mass squared difference and  
reasonably large mixing angle, it  should clearly reveal itself in the oscillometry 
measurements. For a judicious  
neutrino source the "new oscillation length" $L_{14}$ is expected shorter than 3m. Thus  the needed measurements can be 
implemented with a gaseous spherical TPC of modest dimensions with a 
very good energy and position resolution, detecting nuclear recoils following the coherent  neutrino-nucleus elastic scattering. The best candidates for oscillometry, yielding both monochromatic neutrinos as well as antineutrinos, 
are discussed. A sensitivity in the mixing angle $\theta_{14}$,   $\sin^2{(2\theta_{14})}$=0.1 (99{\%}), can be reached after  a few months of data handling.
\end{abstract}

\begin{keyword}
sterile neutrinos, oscillometry,neutral currents, spherical gaseous TPC.


\end{keyword}

\end{frontmatter}
\section{Introduction.}
A recent analysis of the Reactor Neutrino Anomaly (RNA) \cite{RNA11} led to a 
challenging claim that this anomaly can be explained in terms of a new 
fourth neutrino with a  mass
squared difference much larger than one encounters in neutrino oscillations. In fact assuming that the neutrino mass eigenstates are non 
degenerate one finds\cite{RNA11}:
\beq
 \Delta m^2_{24}=|m_2^2-m_4^2|\approx \Delta m^2_{14}=|m_1^2-m_4^2|\ge1.5\mbox{(eV)}^2.
 \eeq
and a mixing angle 
\beq
\sin^2{2 \theta_{14}}=0.17\pm 0.1 (95\%).
\eeq

It is obvious that this new neutrino should contribute to the oscillation 
phenomenon. In the present paper we will assume that the new netrino is sterile, that is it does not participate in weak interaction. Even then, however, it has an effect on neutrino oscillations since it will tend to decrease the elecron neutrino flux. This makes the analysis of oscillation experiments more 
sophisticated. In all the previous experiments the oscillation length is 
much larger than the size of the detector. So one is able to see the effect only 
if the detector is placed in the right distance from the source. It is, 
however, possible to design an experiment with an oscillation length of the 
order of the size of the detector, as it was proposed in \cite{VERGIOM06},\cite{VERNOV10}. This is 
equivalent to many standard experiments done simultaneously.\\
 In a previous paper \cite{VerGiomNov11} we have studied the possibility of 
investigating the low energy neutrino oscillations by the measurement of the electron recoils following neutrino electron scattering.  In the present study we will explore the neutral current interaction to measure nuclear recoils.
The main 
requirements are the same as given previously \cite{VERNOV10} with some modifications pertinent to the neutrino nucleus interaction. More specifically:
 The neutrinos should have  low  energy so that the oscillation 
length is smaller than the size of the detector. At the same time it should sufficiently high so that 
the neutrino-nucleus elastic scattering yields recoils above threshold with a sizable cross section.
A monoenergetic neutrino source is preferred, since it has the advantage that some of the features 
of the oscillation  pattern are not affected as they might be by the averaging over a 
continuous neutrino spectrum. Antineutrino sources with a relatively high energy can also be employed. The "wave form" of the oscillation is not much different than that of the monochromatic source, since only the small portion of the spectrum at the high energy end becomes relevant.
The lifetime of the source should be suitable for the experiment to be 
performed. Clearly a 
compromise has to be made in the selection of the source.

 In this article we will show that, unlike the standard neutrino case, for a sterile neutrino one can observe neutrino diasapperance  oscillations via the neutral current interaction.
Furthermore the aim of this article is to show that the existence of a new fourth 
neutrino can be verified experimentally by the direct measurements 
of the oscillation curves for the monoenergetic neutrino-nucleus elastic 
scattering. It can be done point-by-point within the dimensions of the detector, 
thus providing what we call  neutrino oscillometry \cite{VERNOV10},\cite{VERGIOMNOV}. 
 \section{Neutrino oscillations and neutral current detection}
 Suppose in addition to the three standard neutrinos we have a 4th sterile neutrino. In the neutral current detection all contributing neutrinos have  the same cross section, say $\sigma$. Let us suppose that we initially have electronic neutrinos $\nu_e$. Then we distinguish the following cases:
\begin{itemize}
\item All four neutrinos are active.\\
Then
\beq
\sigma_{\mbox{tot}}=\left (P(\nu_e\rightarrow \nu_e)+P(\nu_e\rightarrow \nu_{\mu})+P(\nu_e\rightarrow \nu_{\tau})+P(\nu_e\rightarrow \nu_4)\right )\sigma,
\eeq
but
\beq
P(\nu_e\rightarrow \nu_e)=1-\left (P(\nu_e\rightarrow \nu_{\mu})+P(\nu_e\rightarrow \nu_{\tau})+P(\nu_e\rightarrow \nu_4)\right )
\label{dis},
\eeq
i.e.
\beq\sigma_{\mbox{tot}}=\sigma,
\eeq
no oscillation is observed.
\item The fourth neutrino is sterile.\\
Then 
\beq
\sigma_{\mbox{tot}}=\left (P(\nu_e\rightarrow \nu_e)+P(\nu_e\rightarrow \nu_{\mu})+P(\nu_e\rightarrow \nu_{\tau})\right )\sigma,
\eeq
i.e. the sterile neutrino does not contribute. Eq. \ref{dis}, however,  is still valid (neutinos are lost from the flux). Thus
\beq
\sigma_{\mbox{tot}}=\left (1-P(\nu_e\rightarrow \nu_4)\right ) \sigma.
\eeq
If, in addition, the new oscillation length is much smaller than the other two, one finds:
\beq
\sigma_{\mbox{tot}}=\left (1-\sin^2{2 \theta_{14}}\sin^2{\pi \frac{L}{L_{14}}} \right )\sigma.
\label{oscsterile}
\eeq
\end{itemize}
It is worth comparing the neutral current situation with neutrino-electron elastic scattering previously considered \cite{VerGiomNov11}. Then for the standard neutrinos we have 
\beq
\sigma_{\mbox{tot}}= P(\nu_e\rightarrow \nu_e) \sigma +\left (P(\nu_e\rightarrow \nu_{\mu})+P(\nu_e\rightarrow \nu_{\tau})\right )\sigma',
\eeq
where $\sigma$ is the $(\nu_e,e)$ cross section, while $\sigma'$  is the cross section for the other two flavors,$(\nu_{\alpha},e)$, $\alpha=\mu,\tau$,. Furthermore
\beq
P(\nu_e\rightarrow \nu_e)=1-\left (P(\nu_e\rightarrow \nu_{\mu})+P(\nu_e\rightarrow \nu_{\tau})\right ).
\eeq
 If, in addition, the new oscillation length is much smaller than the other two, one finds:
\beq
\sigma_{\mbox{tot}}\approx\left (1-\chi(E_{\nu})\sin^2{2 \theta_{13}}\sin^2{\pi \frac{L}{L_{13}}} \right )\sigma,
,\quad \chi(E_{\nu})=\left( 1-\frac{\sigma'}{\sigma}\right ),
\eeq
 i.e. one observes oscillations,  since \cite{VERGIOM06},\cite{VERNOV10} $\sigma'\ne \sigma$.\\
  Furthermore, if  there exists an additional sterile neutrino,  we find that the corresponding oscillation
  cross section in $(\nu,e)$ scattering is still given by Eq. (\ref{oscsterile}). 
\section{The differential and total cross section}
The  elastic neutrino-nucleus scattering due to the neutral current interaction has been previously considered for the detection of sky \cite{VERGIOM06}, \cite{JDVPARIS10} and earth neutrinos \cite{VerAvGiom09} of appreciably higher neutrino energies. It has never been considered in the context of neutrino oscillations, since for standard neutrinos, as we have seen, oscillations in such a channel are not expected.
The differential cross  section for a given neutrino energy $E_{\nu}$ can be cast in the form \cite{VERGIOM06}:
\beq
 \left(\frac{d\sigma}{dT_A}\right)(T_A,E_{\nu})=\frac{G^2_F Am_N}{2 \pi}~(N^2/4) F_{coh}(T_A,E_{\nu}),
 \label{elaswAV1}
\eeq
with
\beq
F_{coh}(T_A,E_{\nu})= F^2(q^2)
  \left ( 1+(1-\frac{T_A}{E_{\nu}})^2
-\frac{Am_NT_A}{E^2_{\nu}} \right),
 \label{elaswAV2}
  \eeq
  where $N$ is the neutron number and $F(q^2)= F(T_A^2+2 A m_N T_A)$ is the nuclear form factor.
  The effect of the nuclear form factor depends on the target.
    Integrating the total cross section of Eq.  \ref{elaswAV1} from  $T_A=T_{th}$ to the maximum allowed by the neutrino energy we obtain the total cross section.  The threshold energy $T_{th}$ depends on the detector.
    Since the favorable neutrino energies are of order 1 MeV, the results are not sensitive to the nuclear form factor. They crucially depend on the detector threshold, since the energy of the recoiling nucleus is quite low. This retardation becomes more severe, in the presence of quenching.

\begin{figure}[!ht]
 \begin{center}
 \subfloat[]
 {
 \rotatebox{90}{\hspace{0.0cm} $Q_{\mbox{\tiny{fac}}}(T_A)\rightarrow $}
\includegraphics[scale=0.5]{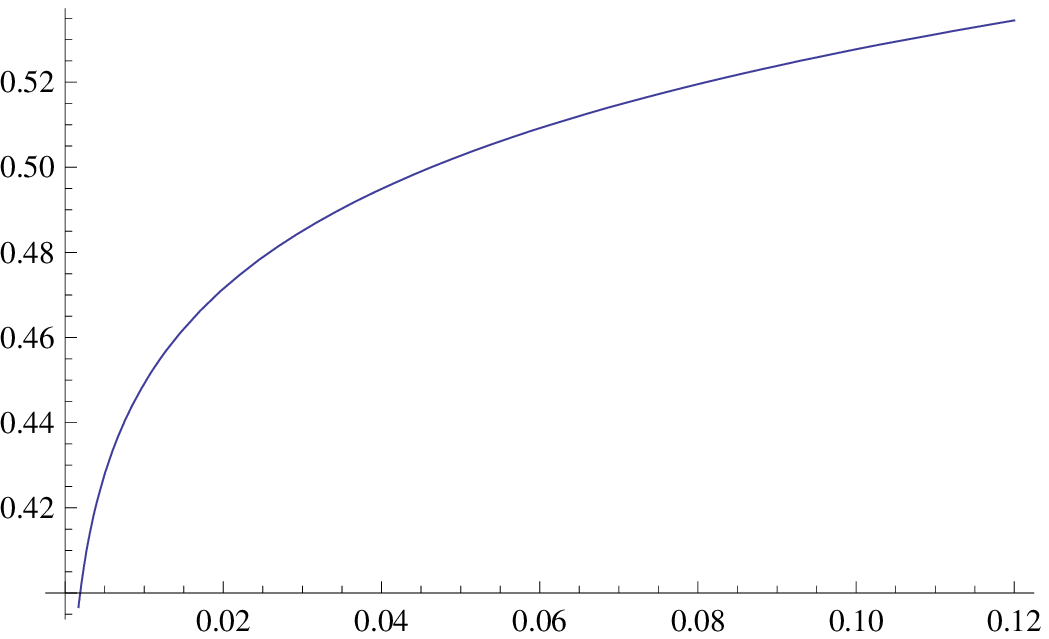}
}
\subfloat[]
{
\rotatebox{90}{\hspace{0.0cm} $T^{\prime}_{th}\rightarrow $ keV}
\includegraphics[scale=0.5]{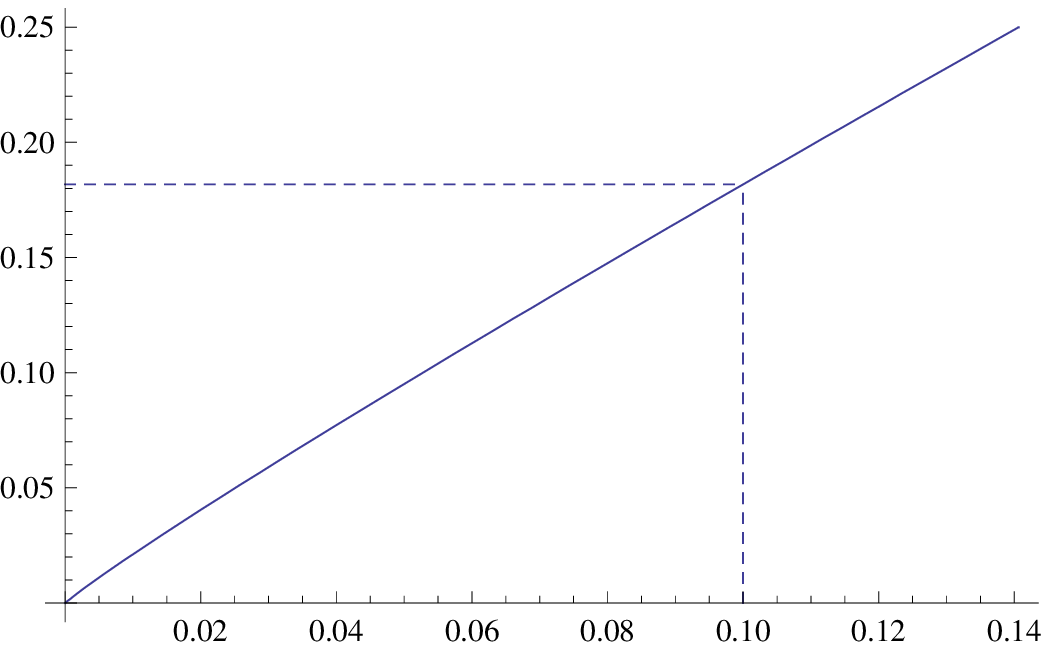}
}\\
\hspace{1.0cm}$T_A\rightarrow$ keV
 \caption{The quenching factor as a function of the recoil energy of interest in the present work (a). Due to quenching the threshold energy for nuclear recoils is shifted upwards from $T_{th}$ to $T^{\prime}_{th}$, e.g. from 0.10 to 0.18 keV (b).}
 \label{fig:quench}
 \end{center}
  \end{figure}

    Furthermore for a real detector the expected nuclear recoil events are quenched, especially at low energies.
The quenching
factor for a given detector  is the ratio of the signal height for a recoil track to that of an electron signal with the same energy. We should not forget that the signal heights depend on the
velocity and how the signals are extracted experimentally. The actual quenching
  factors must be determined experimentally for each target. In the case of NaI the quenching
factor is 0.05, while for Ge and Si it is 0.2-0.3. Thus the measured recoil energy is typically reduced by a factor of about 3 for a Si-detector, when compared with the electron energy. 
 For our purposes it is adequate, to multiply
the energy scale by an recoil energy dependent quenching factor, $ Q_{fac}(T_A)$
  adequately described by the Lindhard theory \cite{LINDHARD}, \cite{SIMON03}.  More specifically in our estimate of $Q_{\mbox{\tiny{fac}}}(T_A)$ we assumed a quenching factor appropriate for a gas target of $^4$He which fits well the data   \cite{SANTOS08} in the energy region of 2 to 50 keV :
\beq
Q_{\mbox{\tiny{fac}}}(T_A)=r_1\left[ \frac{T_A}{1keV}\right]^{r_2},~~r_1\simeq  0.620~~,~~r_2\simeq 0.070
\label{quench1}.
\eeq
The quenching factor as given by Eq.( \ref{quench1}) is exhibited in Fig. \ref{fig:quench}(a) for the energy of interest to us. We will assume that it is the same for all noble gases of interest in the present work.
Due to quenching the threshold energy is shifted upwards, from $T_{th}$ to $T^{\prime}_{th}$ (see Fig.  \ref{fig:quench}(b))

\begin{figure}[!ht]
 \begin{center}
  \subfloat[]
 {
  \rotatebox{90}{\hspace{0.0cm} {$(E_{\nu})_{\mbox{\tiny{min}}}\longrightarrow$MeV}}
\includegraphics[width=2.0in,height=2.0in]{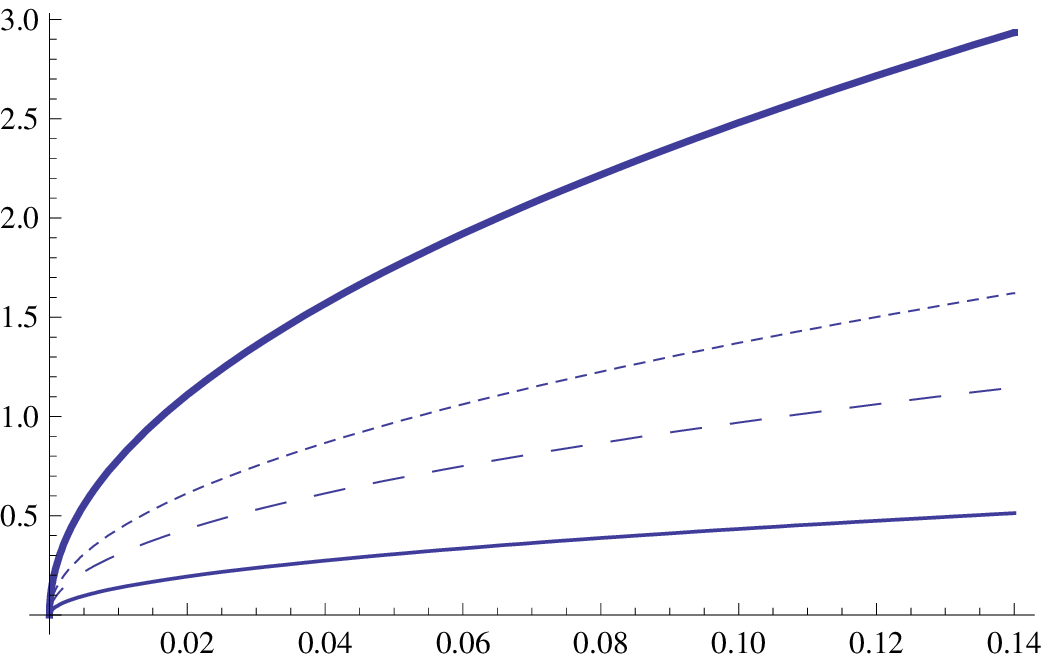}
}
 \subfloat[]
 {
  \rotatebox{90}{\hspace{0.0cm} {$(E_{\nu})_{\mbox{\tiny{min}}}\longrightarrow$MeV}} 
 \includegraphics[width=2.0in,height=2.0in]{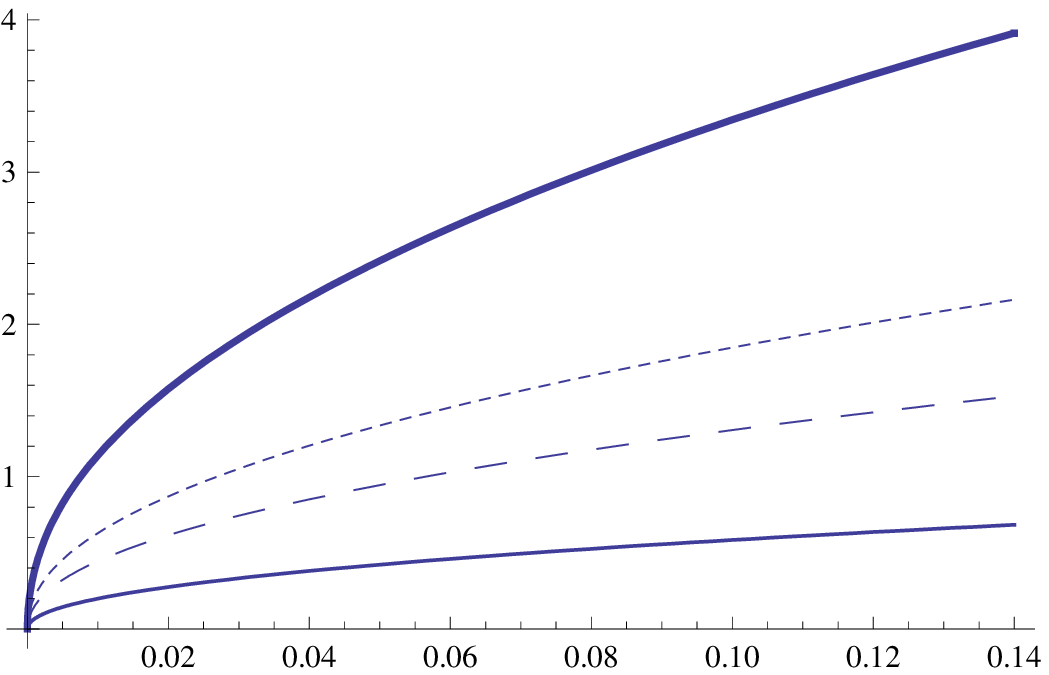}
}\\
\hspace*{-0.0cm} { $ T_{\mbox{\tiny{min}}}\rightarrow$ keV}\\
 \caption{ We show the minimum neutrino energy required as a function of threshold without quenching (a) and with quenching in (b). From top to bottom for the targets of $^{131}$Xe, $^{40}$Ar, $^{20}$Ne and $^4$He. The threshold value is very  crucial, especially for heavy targets.
  } 
 \label{Enumin}
  \end{center}
  \end{figure}

 The minimum neutrino energy required as a function of threshold is presented in Fig. \ref{Enumin}. Furthermore, even if the neutrino energy is above this minimum and the detection is allowed with a coherent  cross section ($\propto N^2$), the heavier the target, the smaller the recoiling energy is and the more effective the retardation due to the threshold becomes.

Once the usual neutrino-nucleus cross sections $\sigma_A(E_{\nu},0) $ are known one can  show that, due to the oscillation, they depend on the distance of the obsevation point from the source. One finds that the neutrino disappearance as seen in the neutrino-nucleus cross section can be cast in the form:
\beq
\sigma_A(E_{\nu},L)=\sigma_A(E_{\nu},0) \left [ 1-\sin^2{2 \theta_{14}}\sin^2{\left ( 3.72451 L\frac{m_e}{E_{\nu}}\right )} \right ].
\label{Eq:Oscsigma}
\eeq
 
  The obtained results crucially depend on the enegy threshold and the quenching factor. Even though the threshold achieved by the gaseous spherical time projection counter (STPC) is impressive, 0.1 keV, since the neutrino recoiling energy is extremely low, in particilar for heavy targets, most of the neutrino sources do not pass the test. Some of those which pass the test are included in table \ref{table1}. From these sources we will consider for further analysis those, which    can be employed at present, namely $^{37}$Ar, $^{51}$Cr, $^{65}$Zn and $^{32}$P.
  \begin{table}[htbp]
\caption{
Proposed candidates for a new neutrino oscillometry at the 
spherical gaseous TPC. 
Tabulated nuclear data have been taken from \cite{AUDI03}, other data have been 
calculated in this work (see the text for details. The mass of the source was assumed to be 0.1 Kg, except for  $^{37}$Ar and  the antineutrino source, $^{32}$P, for which we took 0.0043 kgr each). We note that the oscillation lengths for the new neutrino \cite{RNA11}, \cite{GIULAV10} estimated in this work are much smaller compared to those of all the earlier experiments.       
\label{table1}}
\begin{center}
\begin{tabular}{|c|c|c|c|c|c|}
\hline
\hline
&   &  &  &  \\
Nuclide& 
$T_{1/2}$ \par & 
$E_{\nu }$ & 
$L_{14}$ & 
$N_{\nu }$ \\
& 
(d)& (keV)&
(m)&  
(s$^{-1})$ \\
\hline
$^{37}$Ar& 
35 & 
811& 
1.4& 
$1.8\times 10^{16}$ \\
\hline
$^{51}$Cr& 
27.7 & 
747& 
1.2& 
$4.1\times 10^{17}$ \\
\hline
$^{65}$Zn& 
244 & 
1343& 
2.2& 
$3.0\times 10^{16}$ \\
\hline
$^{59}$Ni& 
$2.8\times 10^7$ & 
1065& 
1.8& 
$1.1\times 10^{14}$ \\
\hline
$^{113}$Sn& 
116 & 
617& 
1.0& 
$3.7\times 10^{16}$ \\

\hline
$^{32}$P& 
14.3 & 
continuum& 
$\approx 2.5$& 
$5.0\times 10^{16}$ \\
\hline
\hline
\end{tabular}
\end{center}
\end{table}
 \section {Results  with monochromatic sources}
 We will begin with the cross sections and proceed to the event rates.
 \subsection{Coherent cross sections}
   For a light target like He, a variety of sources pass the test, but one loses the benefit of large coherence. In comparing these results with those of neutrino-electron scattering we should keep in mind that the present cross sections explicitly contain the $N^2$ term due to  coherence, in the case of  neutrino-electron scattering the enhancement $Z$ is not included in the cross section, but it has been contained in the electron densiity, which is $Z$ times the number of nuclei per unit volume. 
   
\begin{figure}[!ht]
 \begin{center}
  \subfloat[]
 {
  \rotatebox{90}{\hspace{0.0cm} {$\sigma_A(x,L)/((G_F m_e)^2/(2 \pi))\longrightarrow$}}
\includegraphics[width=2.3in,height=2.0in]{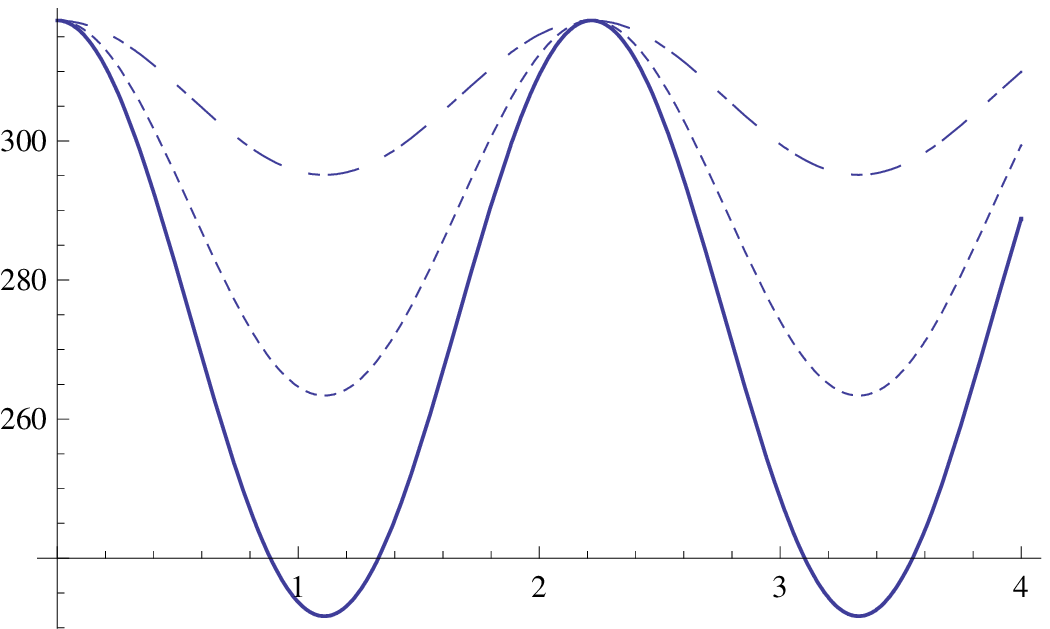}
}
 \subfloat[]
 {
  \rotatebox{90}{\hspace{0.0cm} {$\sigma_A(x,L)/((G_F m_e)^2/(2 \pi))\longrightarrow$}}
 \includegraphics[width=2.3in,height=2.0in]{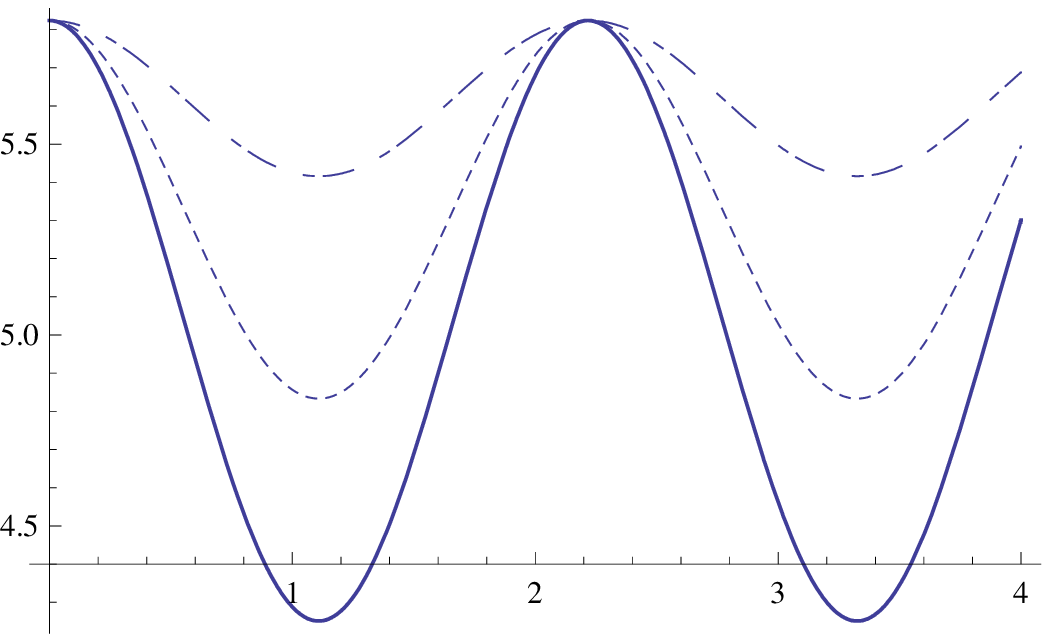}
}\\
\hspace*{-0.0cm} { $L \rightarrow$ meters}\\
 \caption{ In panel (a) we show the total cross section in units of $(G_Fm_e)^2/(2\pi)=2.29 \times 10^{-49}$m$^2$  as a function of the distance of the  recoiling nucleus from the source in meters assuming a threshold of 0.1 keV. In panel (b) we show the same quantity in the presence of quenching. In both panels the solid, dotted and dotted-dashed  curves correspond to $\sin^2{2 \theta_{14}}=0.27,0.17,0.07$ respectively. The results were obtained for a Ne target with the $^{65}$Zn source.
  } 
 \label{sNeZn}
  \end{center}
  \end{figure}

\begin{figure}[!ht]
 \begin{center}
  \subfloat[]
 {
  \rotatebox{90}{\hspace{0.0cm} {$\sigma_A(x,L)/((G_F m_e)^2/(2 \pi))\longrightarrow$}}
\includegraphics[width=2.3in,height=2.0in]{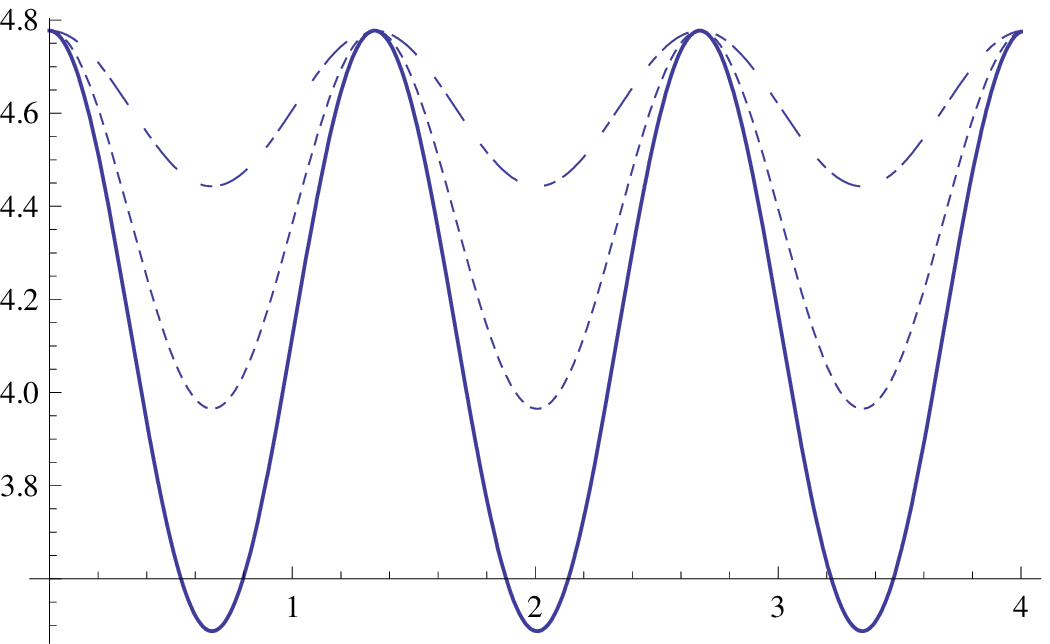}
}
 \subfloat[]
 {
  \rotatebox{90}{\hspace{0.0cm} {$\sigma_A(x,L)/((G_F m_e)^2/(2 \pi))\longrightarrow$}}
 \includegraphics[width=2.3in,height=2.0in]{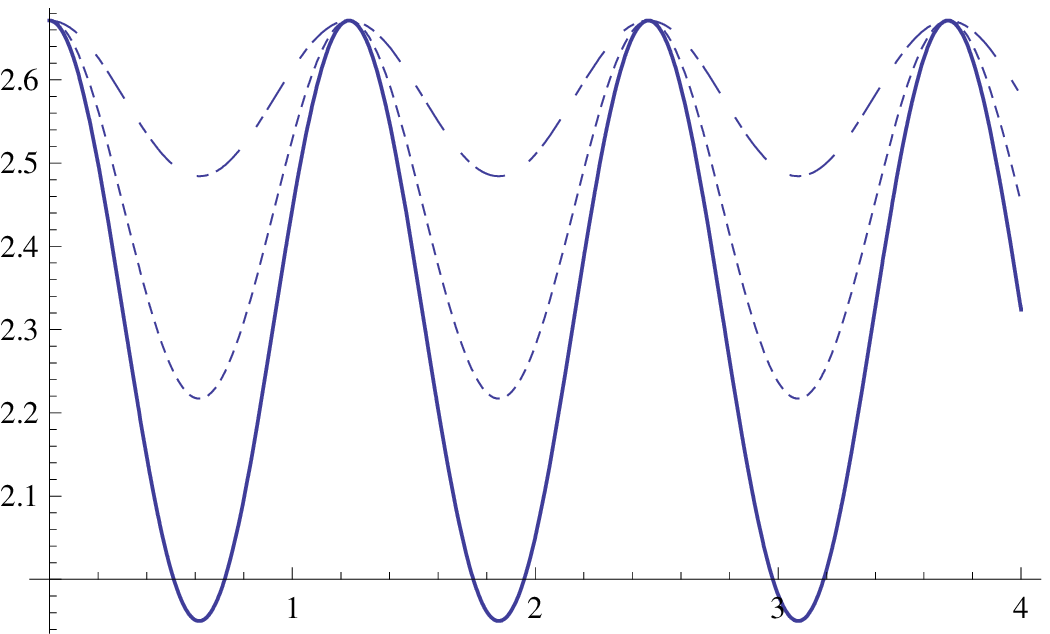}
}
\\
\hspace*{-0.0cm} { $L \rightarrow$ meters}\\
\caption{In panel (a) we show the total cross section in units of  $(G_Fm_e)^2/(2\pi)=2.29\times 10^{-49}$m$^2$ as a function of the distance of the  recoiling nucleus from the source in meters assuming a threshold of 0.1 keV. The results were obtained for a He target with the $^{37}$Ar source. In panel (b) we show the same quantity for a $^{51}$Cr source. In both panels the effect of quenching was included.  The solid, dotted and dotted-dashed  curves correspond to $\sin^2{2 \theta_{14}}=0.27,0.17,0.07$ respectively.}
 \label{sHeAr}
  \end{center}
  \end{figure}

   For the sources and the targets that pass the test we present our results in Figs \ref{sNeZn}-\ref{sHeAr}.
  
  \subsection{The event rate}
  We will consider a spherical detector with the source at the origin and will 
assume that the volume of the source is much smaller than the volume of the 
detector. 
The  event rate $dI$ between $L$ and $L+dL$ is given by:
\beq
dI=N_{\nu} n_A \frac{4 \pi L^2dL}{4 \pi L^2} \sigma(L,x)=N_{\nu} n_A dL \sigma(L,x),
\eeq
where $N_{\nu}$ is the neutrino intensity and $n_A$ is the number of the target nuclei:
\beq
n_A=\frac{P}{k T_0},
\eeq
where $P$ is its pressure and $T_0$ its temperature. Thus
\beq
 \frac{dI}{dL}=N_{\nu} n_A \sigma(L,x)
 \label{eventsph}
\eeq
%
%
%
%
%
%
or
 \beq
  R_0\frac{dI}{dL}=\Lambda\tilde{\sigma}(L,x),
\eeq
where
\beq
\Lambda=\frac{G^2_F m^2_e}{2 \pi} R_0 N_{\nu} n_A 
\eeq
 $R_0$ the radius of the target and $\tilde{\sigma}(L,x)$ is the neutrino-nucleus cross section in units of 
$(G_Fm_e)^2/(2\pi)=2.29\times 10^{-49}$m$^2$.
The total number of events per unit length after running for time $t_r$ will be given by
\beq
\frac{d N}{d L}=\Lambda_1 \tilde{\sigma}(L,x)\left (1-e^{-t_r/\tau}\right ),\quad \Lambda_1=\frac{G^2_F m^2_e}{2 \pi} \tau N_{\nu} n_A 
\label{Eq:Nevents}
\eeq
where $\tau$ is the lifetime of the source.
Integrating Eq. (\ref{Eq:Nevents}) over $L$ from 0 to $R_0$ we obtain the total number of events, which can be cast in the form:
\beq
N=A+B\sin^2{2 \theta_{14}}.
\label{Eq:AB}
\eeq
The parameters $A$ and $B$ for some cases of interest are presented in table \ref{table2}. We have included here the relevant  results even for the pairs source-target not discussed further in the present work in the context of oscillometry\footnote{ Due to lack of space the expected oscillation patterns are not provided in this paper. They can be obtained by communicating directly with the authors}, just to give an estimate on the uncertainties expected in the extraction of $\sin^2{2 \theta_{14}}$  from the total number of events
\begin{table}[htbp]
\caption{ The parameters $A$ and $B$ entering the total event rate (for their definition see the text, Eq. (\ref{Eq:AB})). The source $^{205}$Bi is included to show the importance   of the size of the energy of a monochromatic source, even though it is, at present, unrealistic (in the sense of production) to employ it.
\label{table2}}
\begin{center}
\begin{tabular}{|c|c|c|c|c|c|}
\hline
\hline
&   &  &  &  \\
target-& 
$A$ (no quenching) & 
$B$ (no quenching)& 
$A$ (quenching)& 
$B$ (quenching) \\
 source&& & &  \\
\hline
$^{40}$Ar-$\,^{32}$P& $2.4\times 10^2 $& $-1.2\times 10^2$& &  \\
$^{40}$Ar-$\,^{205}$Bi& $1.4\times 10^4 $& $-6.6\times 10^3$& $4.2\times 10^2 $& $-1.8\times 10^{2}$ \\
\hline
$^{20}$Ne-$\,^{32}$P& $8.8\times 10^2 $& $-4.6\times 10^2$& $1.0\times 10^2 $& $-5.4\times 10$ \\
$^{20}$Ne-$\,^{65}$Zn& $2.9\times 10^4 $& $-1.6\times 10^4$& $5.3\times 10^2 $& $-2.8\times 10^2$ \\
\hline
$^{20}$Ne-$\,^{205}$Bi& $7.2\times 10^3 $& $-3.3\times 10^3$& $3.8\times 10^3 $& $-1.7\times 10^{3}$ \\
\hline
$^{4}$He-$\,^{37}$Ar& $7.8\times 10 $& $-3.9\times 10$& $3.6\times 10 $& $-1.8\times 10$ \\
\hline
$^{4}$He-$\,^{51}$Cr& $8.7\times 10^2 $& $-4.1\times 10^2$& $3.1\times 10^2 $& $-1.5\times 10^{2}$ \\
\hline
$^{4}$He-$\,^{65}$Zn& $4.0\times 10^3 $& $-2.1\times 10^3$& $3.3\times 10^3 $& $-1.8\times 10^{3}$ \\
\hline
$^{4}$He-$\,^{205}$Bi& $4.6\times 10^2 $& $-2.0\times 10^2$& $4.3\times 10^2 $& $-1.9\times 10^{2}$ \\
\hline
\hline
\end{tabular}
\end{center}
\end{table}

The goal of the experiment is to scan the monoenergetic neutrino nucleus 
elastic scattering events by measuring the nuclear recoils  as a function of 
distance from the neutrino source prepared in advance at the reactor/s. This 
scan means point-by-point determination of scattering events along the 
detector dimensions within its position resolution.
These events can be observed as a smooth curve, which 
reproduces the neutrino disappearance probability.
It is worthwhile to note again that the 
oscillometry is suitable for monoenergetic neutrino, since it deals with a 
single oscillation length or $L_{14}$(see table \ref{table1}). This is obviously not a case 
for antineutrino, since, in this instance, one extracts only an effective 
oscillation length. This could be a serious problem in neutrino electron scattering. In the case of nuclear recoils, as we will see below, with a judicious choice of the target-source pair, not much  information is lost due to the folding, 
since only a narrow band in the high energy tail of the continuous neutrino energy spectrum contributes to nuclear recoils, even though the assumed threshold for noble gas targets is quite low, 0.1 keV.

Table \ref{table1} clearly shows that the oscillation lengths for a new neutrino 
proposed in \cite{RNA11} are much smaller compared to those previously considered \cite{VERGIOMNOV} in connection with $\theta_{13}$. They can thus  be  directly measured within 
the dimensions of detector of reasonable sizes. One of the very promising 
options could be the Spherical Time Projection Counter (STPC)  proposed 
in \cite{VERGIOM06}. If necessary, a spherical Micromegas based on the micro-Bulk  
 technology \cite{ADRIAm10},
 which will be developed in the near future, can be employed in the STPC.  
 A thin 50
 micron polyamide foil will be used as bulk material  to fabricate the
 detector structure. This detector provides an excellent energy  
 resolution, can
 reach high gains at high gas pressure (up to 10 Atm) and has the advantage that its  
 radioactivity
 level \cite{CEBRIAN10} should  fulfill the  requirements of the proposed  
 experiment.

In this spherical chamber with a modest radius, assumed to be 4m, the 
neutrino source can be situated in the center of the sphere and the 
detector for recoils is also placed around the source in the smaller sphere with radius 
$r \approx 0.5$ m. The chamber  outside this small sphere  is filled with a 
gas (a noble gas such as Ar, Ne, or He (preferably Ar), if the neutrino source is of sufficiently high energy). In the present work we assumed the gas is at room temperature under a pressure of 10 Atm.

 The nuclear recoils  are guided by a Micromegas-detector \cite{Giomataris},\cite{GIOMVER08}. Such type of device has an 
advantage in precise position determination (better than 0.1 m) and in 
detection of very low nuclear recoils in 4$\pi$-geometry (down to a few 
hundreds of eV, that well suits  the nuclides from Table \ref{table1}).
 
\begin{figure}[!ht]
 \begin{center}
  \subfloat[]
 {
   \rotatebox{90}{\hspace{0.0cm} {$dN/dL\longrightarrow$m$^{-1}$}}
\includegraphics[width=2.3in,height=2.0in]{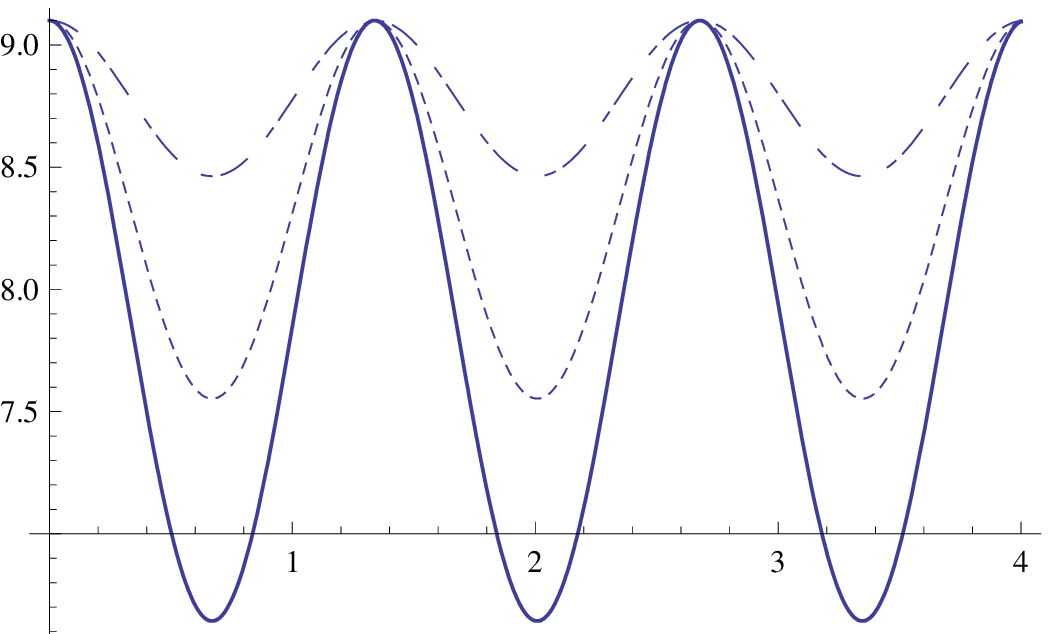}
}
 \subfloat[]
 {
   \rotatebox{90}{\hspace{0.0cm} {$dN/dL\longrightarrow$m$^{-1}$}}
 \includegraphics[width=2.3in,height=2.0in]{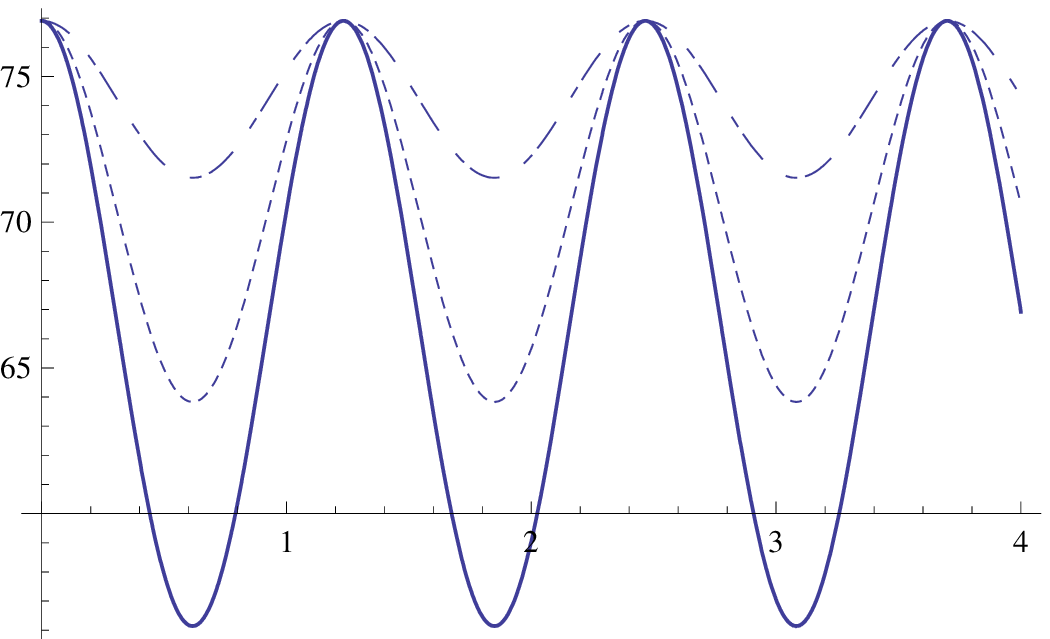}
}\\
\hspace*{-0.0cm} { $L \rightarrow$ meters}\\
  \caption{ In panel (a) we show the number of events as a function of the distance of the  recoiling nucleus from the source in meters assuming a threshold of 0.1 keV in the presence of quenching. The results were obtained for a He target with the $^{37}$Ar source, i.e. $\Lambda_1=1.9$m$^{-1}$. In panel (b) we show the same quantity with the $^{51}$Cr source, i.e. $\Lambda_1=28.8$m$^{-1}$. In both panels the target gas was enclosed in a sphere of 4m radius, at temperature T=300 $^0$K and under a pressure of 10 Atm. In both panels the solid, dotted and dotted-dashed  curves correspond to $\sin^2{2 \theta_{14}}=0.27,0.17,0.07$ respectively. .
  } 
 \label{qabsHeArHeCr}
  \end{center}
  \end{figure}

\begin{figure}[!ht]
 \begin{center}
  \subfloat[]
 {
   \rotatebox{90}{\hspace{0.0cm} {$dN/dL\longrightarrow$m$^{-1}$}}
\includegraphics[width=2.3in,height=2.0in]{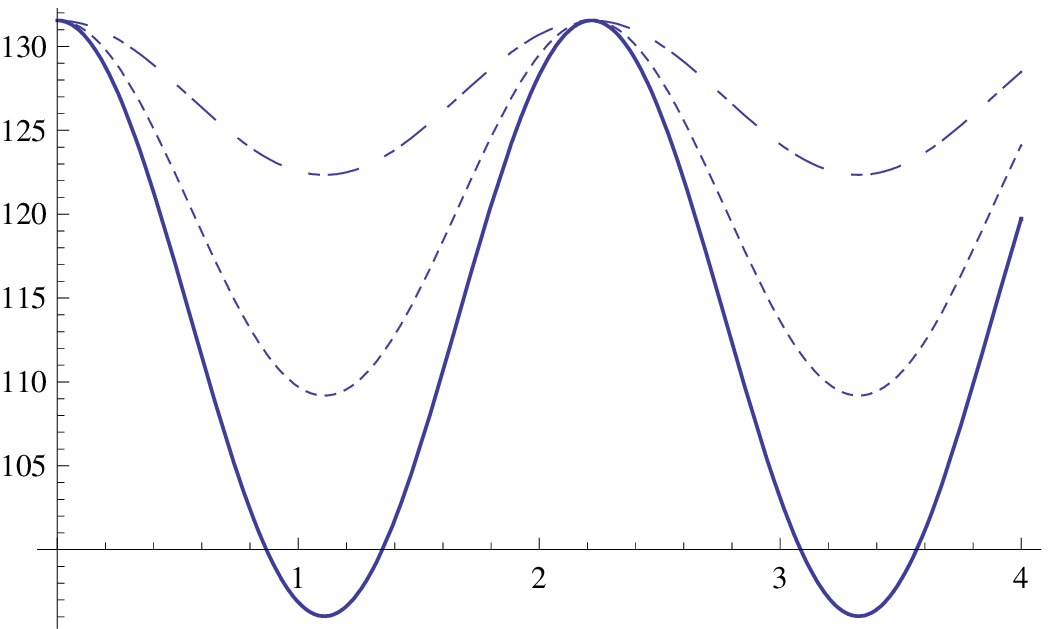}
}
 \subfloat[]
 {
   \rotatebox{90}{\hspace{0.0cm} {$dN/dL\longrightarrow$m$^{-1}$}}
 \includegraphics[width=2.3in,height=2.0in]{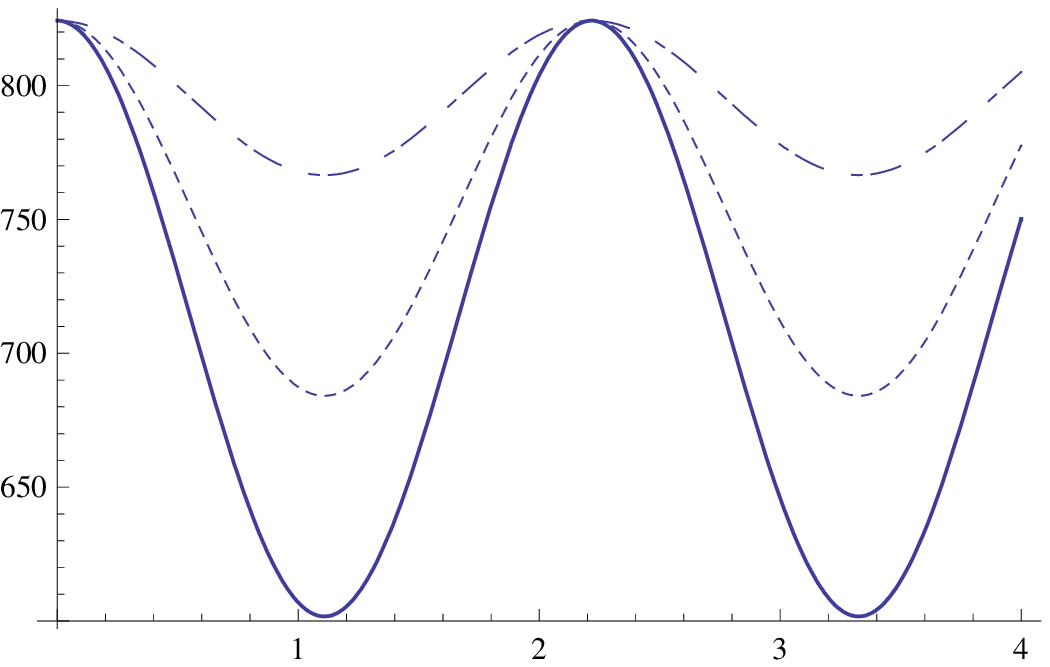}
}\\
\hspace*{-0.0cm} { $L \rightarrow$ meters}\\
 \caption{ The same as in Fig. \ref{qabsHeArHeCr}  for a Ne target (a) and He target (b) with the $^{65}$Zn source, i.e $\Lambda_1=22.6$m$^{-1}$. } 
 \label{qabsNeZnHeZn}
  \end{center}
  \end{figure}

The results obtained in the presence of quenching are presented in Figs \ref{qabsHeArHeCr}-\ref{qabsNeZnHeZn}

  \section{Antineutrino Sources}
\begin{figure}[!ht]
 \begin{center}
 \subfloat[]
 {
\includegraphics[width=2.3in,height=2.0in] {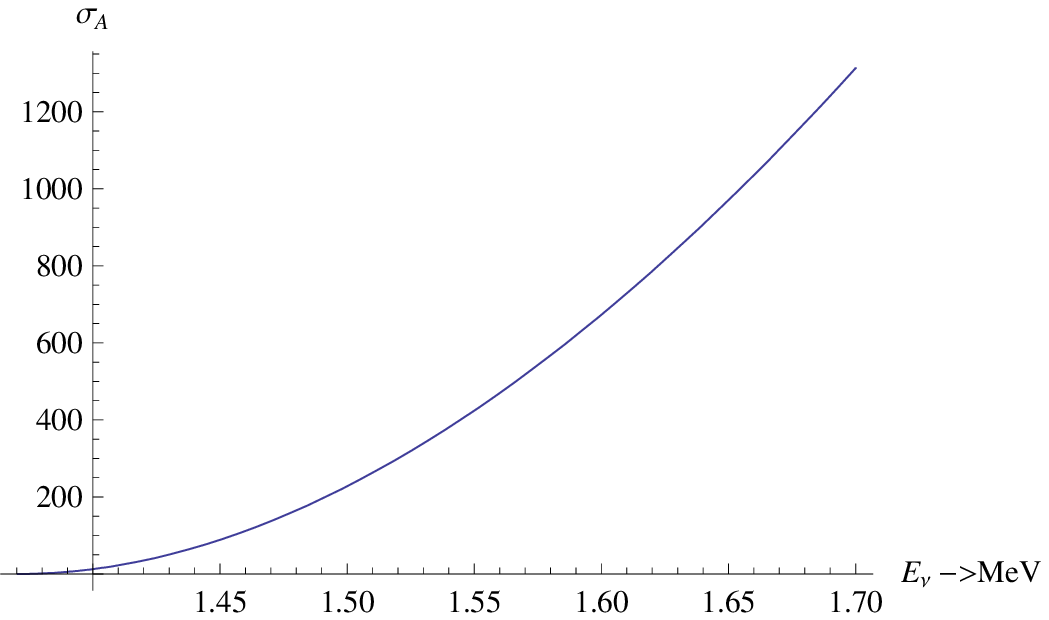}
}
 \subfloat[]
 {
\includegraphics[width=2.3in,height=2.0in] {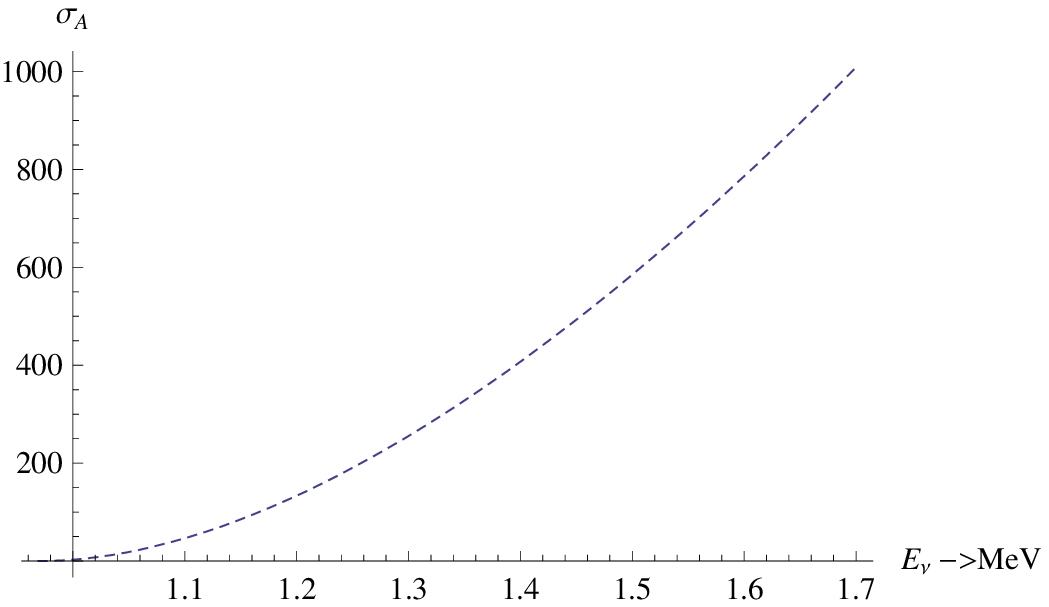}
}
 \caption{ The non oscillating part of the antineutrino cross section in units of $(G_Fm_e)^2/(2\pi)=2.29\times 10^{-49}$m$^2$  as a function of the energy in MeV for a target $^{40}$Ar (a) and  $^{20}$Ne (b), assuming a threshold of 0.1 keV.}   
 \label{antisigma40}
  \end{center}
  \end{figure}

\begin{figure}[!ht]
 \begin{center}
\includegraphics[width=4.3in,height=3.0in] {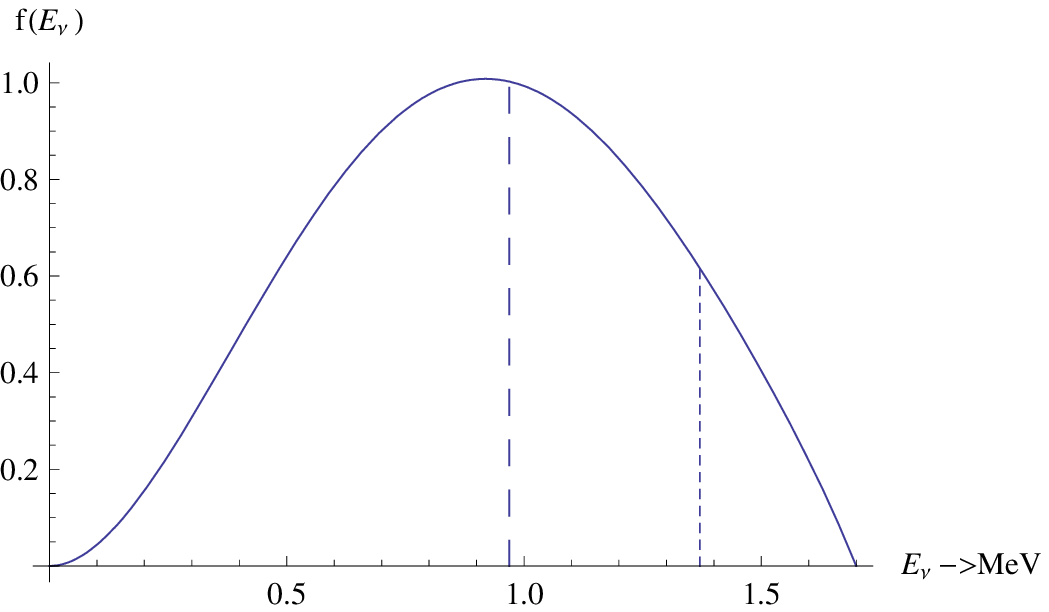}
 \caption{ The normalized antineutrino spectrum following the beta decay of $^{32}$P. The vertical line indicates the space on its right allowed for  $^{40}$Ar (dotted line) and $^{20}$Ne  (dashed line) targets, assuming a threshold of 0.1 keV.}   
 \label{antispec32}
  \end{center}
  \end{figure}

\begin{figure}[!ht]
 \begin{center}
 \subfloat[]
 {
 \rotatebox{90}{\hspace{0.0cm} {$\sigma_A(L)/((G_F m_e)^2/(2 \pi))\longrightarrow$}}
\includegraphics[width=2.0in,height=2.0in] {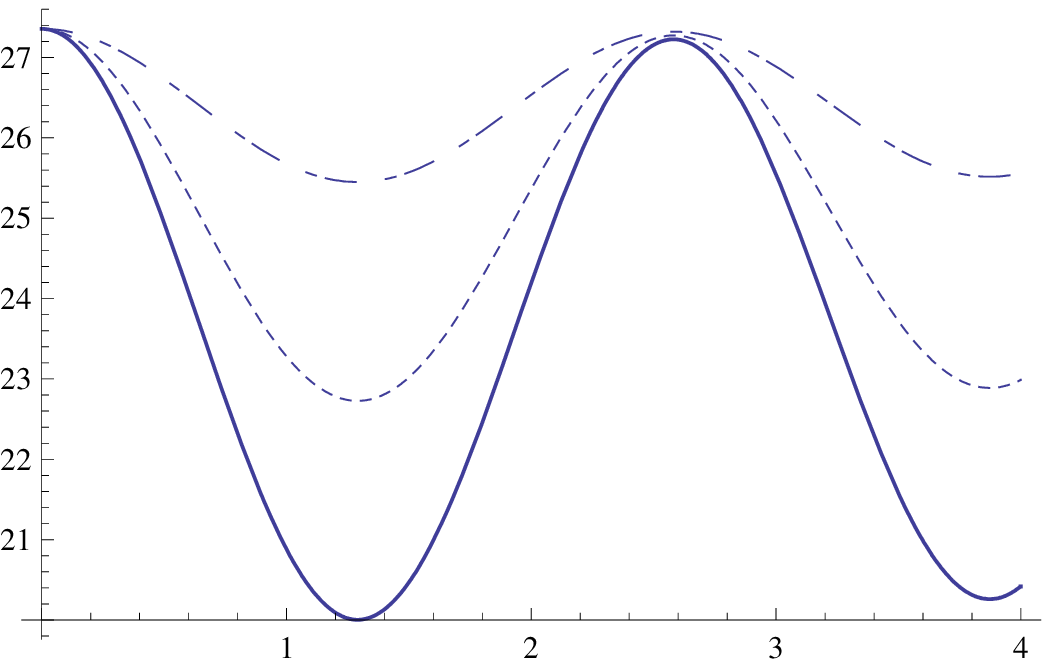}
}
\subfloat[]
 {
 \rotatebox{90}{\hspace{0.0cm} {$\sigma_A(L)/((G_F m_e)^2/(2 \pi))\longrightarrow$}}
\includegraphics[width=2.0in,height=2.0in] {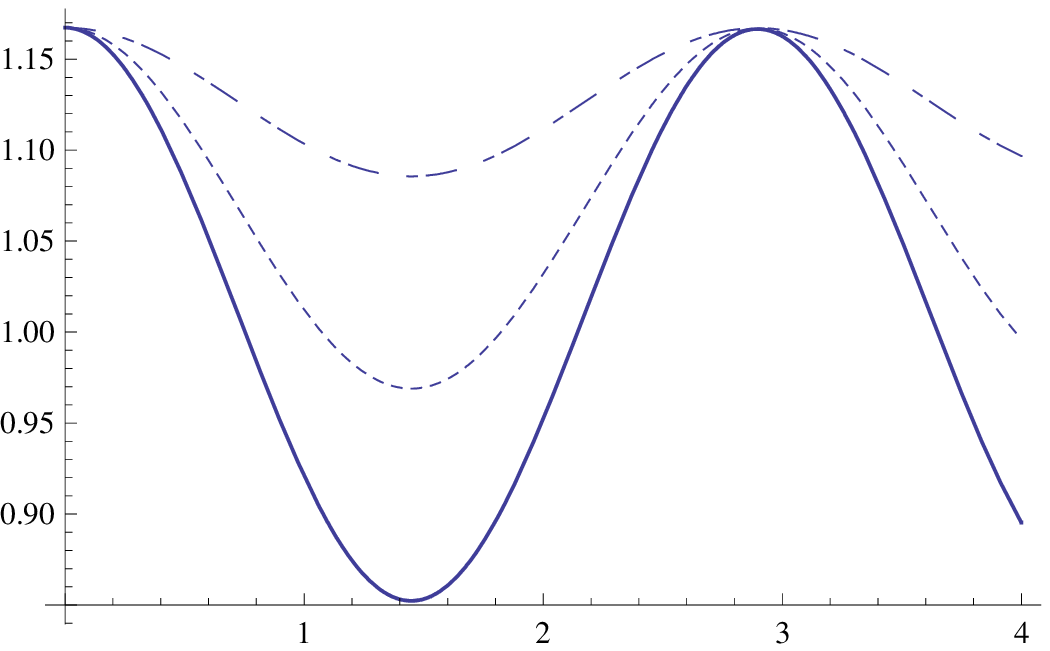}
}
\hspace*{-0.0cm} { $L \rightarrow$ meters}\\
 \caption{ The antineutrino cross section in units of $(G_Fm_e)^2/(2\pi)=2.29\times 10^{-49}$m$^2$, folded with the spectrum of Fig. \ref{antispec32}, as a function of $L$ in m for a target $^{40}$Ar, assuming an energy threshold of 0.1 keV and $^{32}$P source, without the quenching factor (a) and including quenching (b).The solid, dotted and dotted-dashed  curves correspond to $\sin^2{2 \theta_{14}}=0.27,0.17,0.07$ respectively. 
 }   
 \label{fig:antnuoscP32Ar}
  \end{center}
  \end{figure}

\begin{figure}[!ht]
 \begin{center}
 \subfloat[]
 {
 \rotatebox{90}{\hspace{0.0cm} {$\sigma_A(L)/((G_F m_e)^2/(2 \pi))\longrightarrow$}}
\includegraphics[width=2.5in,height=2.0in] {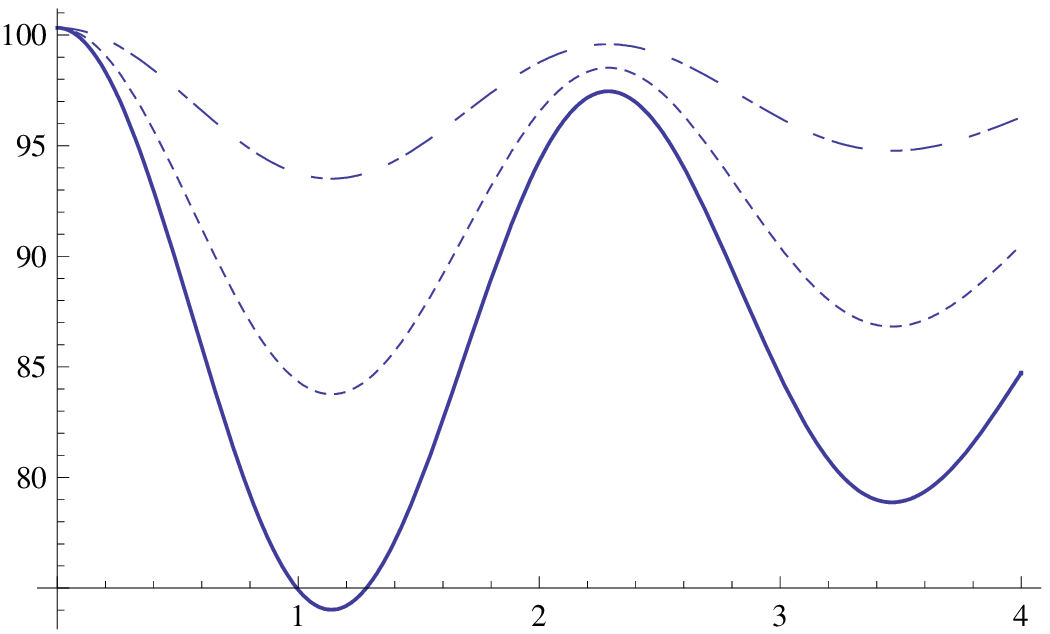}
}
\subfloat[]
{
 \rotatebox{90}{\hspace{0.0cm} {$\sigma_A(L)/((G_F m_e)^2/(2 \pi))\longrightarrow$}}
\includegraphics[width=2.5in,height=2.0in] {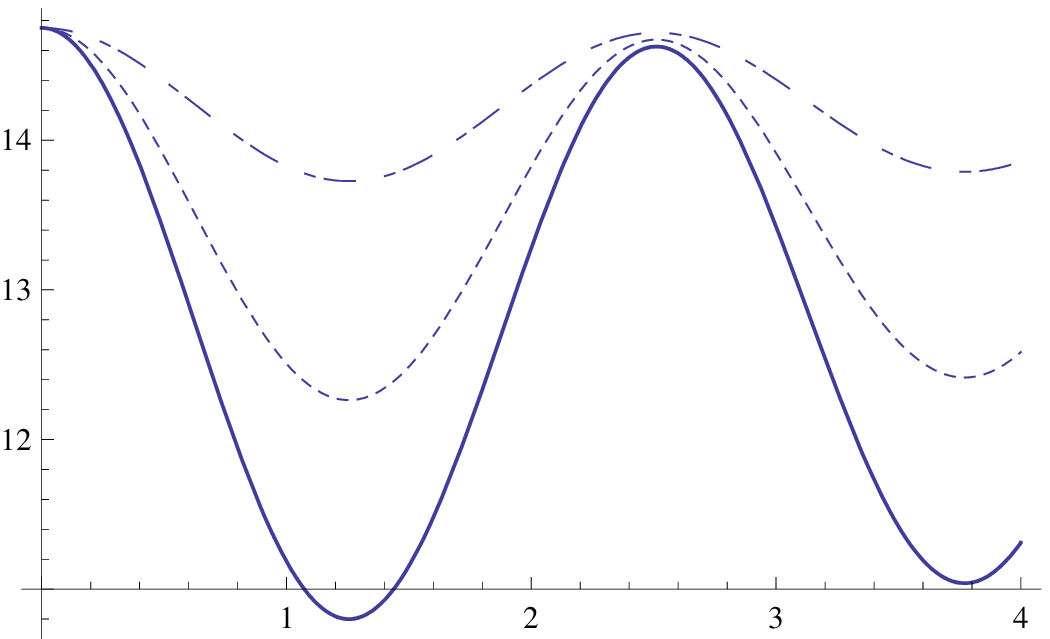}
}\\
\hspace*{-0.0cm} { $L \rightarrow$ meters}\\
 \caption{ The same as in Fig. \ref{fig:antnuoscP32Ar} for the target $^{20}$Ne and the $^{32}$P source , assuming a threshold of 0.1 keV, without the quenching factor (a) and including quenching (b). At the position of the second peak,  we see a  distortion of the "wave form", which, without quenching, is bigger due to the fact that the allowed neutrino energy range is, in this case, wider.}
 \label{fig:antnuoscP32Ne}
  \end{center}
  \end{figure}

\begin{figure}[!ht]
 \begin{center}
  \subfloat[]
 {
  \rotatebox{90}{\hspace{1.0cm} {$dN/dL\longrightarrow$m$^{-1}$}}
\includegraphics[width=2.3in,height=2.0in]{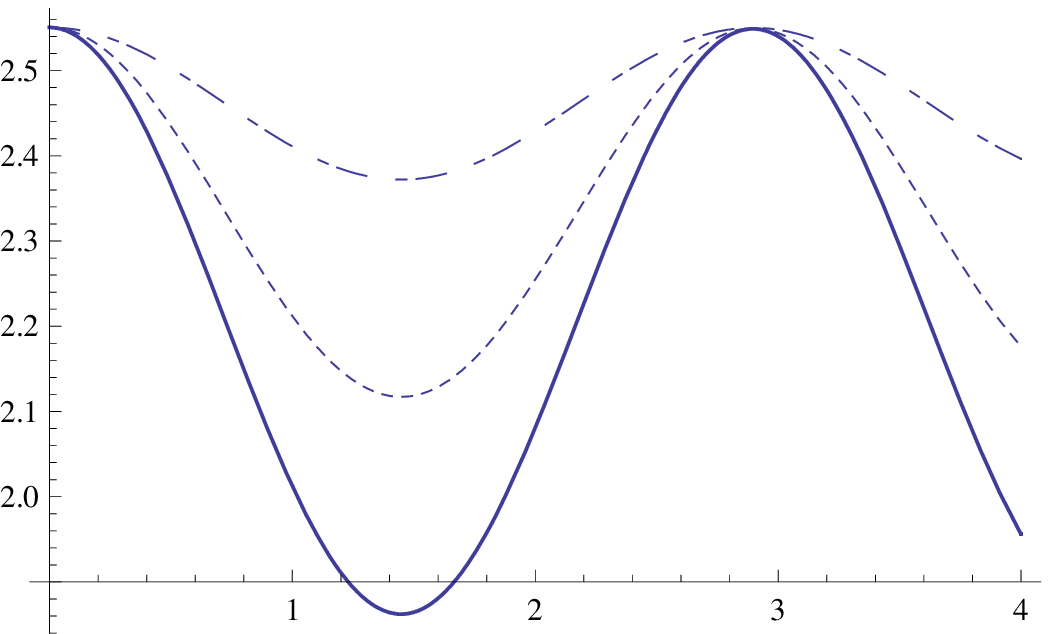}
}
 \subfloat[]
 {
   \rotatebox{90}{\hspace{1.0cm} {$dN/dL\longrightarrow$m$^{-1}$}}
 \includegraphics[width=2.3in,height=2.0in]{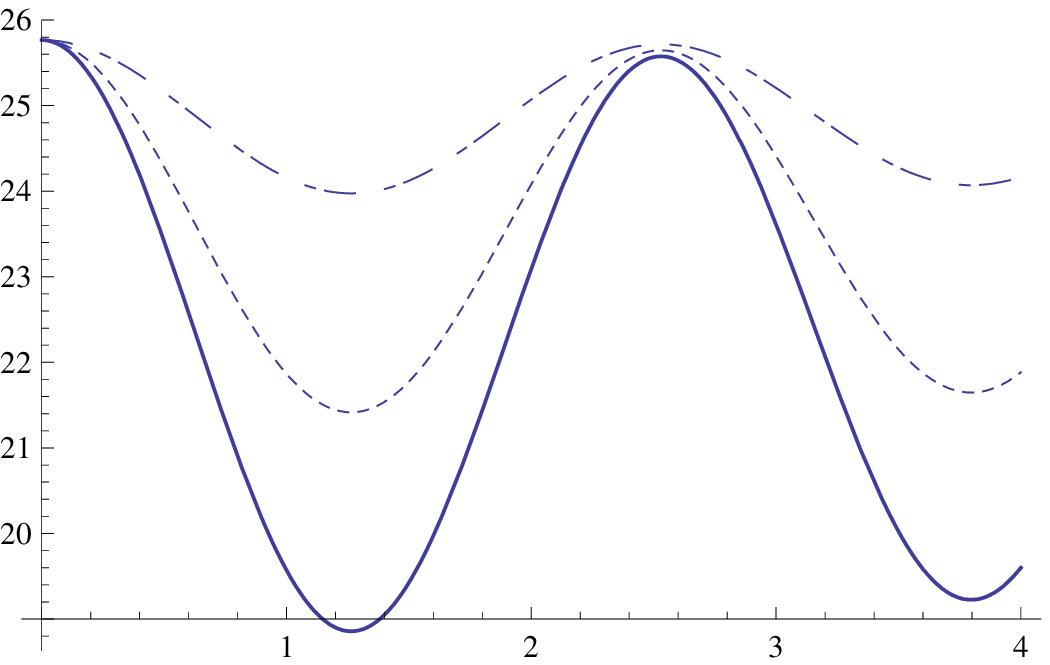}
}\\
\hspace*{-0.0cm} { $L \rightarrow$ meters}\\
 \caption{ In panel (a) we show the number of events for an $^{40}$Ar target as a function of the distance of the  recoiling nucleus from the source $^{32}$P in meters, assuming a threshold of 0.1 keV. In panel (b) we show the same quantity for the $^{20}$Ne target. In both panels the effect of quenching has been included and the target gas was enclosed in a sphere of 4m radius, at temperature T=300 $^0$K and under a pressure of 10 Atm. For the $^{32}$P source we used $\Lambda_1=2.2$m$^{-1}$.  Otherwise the notation is the same as in Fig. \ref{fig:antnuoscP32Ar}
  } 
 \label{absqArNePo}
  \end{center}
  \end{figure}

  The monochromatic neutrino sources considered above, unfortunately, have the disadvantage that the neutrino energy is lower than required to meet the experimental requirements for a neutral current (NC) detector with high neutron number.  Thus one may have to resort to antineutrino sources paying the price of distortion of the form of the oscillation due to the integration over the antineutrino spectrum.
  The NC cross section for an $^{40}$Ar target with a threshold of 0.1 keV as a function of neutrino energy is shown in Fig. \ref{antisigma40}. This cross section must be folded with the energy spectrum of the source.
   From an experimental point of view  $^{32}$P is the best antineutrino source (see table \ref{table1}). In principle, even though it has not been included in table \ref{table1},  $^{90}$Sr can also be used as an antineutrino source. In the case of $^{32}$P 
   the normalized spectrum is exhibited in Fig. \ref{antispec32}. One sees that, due to theshold effects only a portion of the spectrum can be exploited.
 Since the regions of large cross section have a small probability in the spectrum, the integrated cross section is not very large. In fact folding Eq. \ref{Eq:Oscsigma} we obtain the results shown in Figs \ref{fig:antnuoscP32Ar}-\ref{fig:antnuoscP32Ne}.
  From the thus obtained cross sections proceeding as above we obtain the differential number of events $dN/dL$ exhibited in Fig. \ref{absqArNePo}.
  Integrating  over $L$ the results of Fig. \ref{absqArNePo}, and the corresponding one without quenching, not shown here, we obtain  the parameters $A$ and $B$ shown in table \ref{table2}.
\section{Discussion}
We have seen that neutrino oscillometry will lead to a direct 
observation of the fourth sterile neutrino in electron neutrino disapperarnce experiments, if such a neutrino exists. The calculations and analysis 
shows that the gaseous STPC is a powerful tool 
for identification of a new neutrino in two ways i) by observing the electron recoils in neutrino-electron scattering , recently discussed by us \cite{VerGiomNov11}, and ii) by measuring the nuclear recoils in the coherent neutrino-nucleus scattering via the neutral current interaction, discussed in this work. The latter method can utilize the wide experience obtained in other experiments attempting to measure nuclear recoils, like dark matter, (quenching factors, background minimization etc). 

We have seen that, since the expected mass-squared difference for this neutrino is rather high, the 
corresponding oscillation length is going to be sufficiently  small for 1 MeV neutrino energy, so that it can 
be fitted into the dimensions of a spherical detector with the radius of a 
few meters. The neutrino oscillometry can be implemented in this detector 
with the use of the intense neutrino sources which can be 
placed at the origin of the sphere. The gaseous STPC with the Micromegas 
detection has a big advantage in the 4$\pi$-geometry and in very good position 
resolution (better than 0.1 m) with a very low energy threshold ($\approx 
$ 100 eV).

The most promising candidates for oscillometry have been 
considered. These are the monochromatic sources $^{37}$Ar,  $^{51}$Cr and $^{65}$Zn 
 as well as $^{32}$P for an antineutrino source.
As an example, for the $^{65}$Zn source the sensitivity to the mixing angle 
$\theta_{14}$ is estimated as $\sin^2{(2\theta_{14})}$ = 0.1 with the 99{\%} of 
confidence, arising from the total number of events collected during only a few months of data handling. The observation of the expected characteristic oscillometry curve will provide a much more precise information on the oscillation length and thus constitute a definite manifestation of the existence of a new type of neutrino as
very recently proposed by the analysis of the reactor antineutrino anomaly.

Unfortunately, however, one cannot fully benefit from the large size of the coherent cross section for large neutron number $N$ of the target. The reason is that larger $N$ implies larger nuclear mass number $A$, i.e. lower recoil energy. Thus, even though the gaseous STPC detector has already achieved the impressive threshold of 0.1 keV, this is still high enough to reduce the cross sections for neutrinos in the MeV range for heavy targets or even exclude these targets.
 It is clear that the accuracy of the extraction of the above parameters will substantially increase only if the threshold can be further reduced, since it does not seam feasible to utilize a monochromatic  neutrino source with energy  higher than that of 1343 keV associated with $^{65}$Zn.

  \end{document}